\documentclass[10pt]{iopart} 
\usepackage{graphicx}
\usepackage{enumitem}
 
\usepackage{iopams}  
\newcommand{\aap}{{\it Astron. Astrophys.}}
\newcommand{\apj}{{\it Astrophys. J.}}
\newcommand{\apjl}{{\it Astrophys. J. Lett.}}

\newcommand{\mnras}{{\it Mon. Not. Roy. Astron. Soc.}}
\newcommand{\ppcf}{{\it Plasma Phys. Control. Fus.}}
\newcommand{\pre}{\it Phys. Rev. E}
\newcommand{\solphys}{{\it Solar Phys.}}
\newcommand{\ssr}{{\it Space Sci. Rev.}}
\newcommand{\araa}{{\it Annu. Rev. Astron. Astrophys.}}
\newcommand{\apss}{{\it Astrophys. Space Sci.}}

\newcommand{\pder}[2]{ \frac{\partial #1}{\partial #2} }
\newcommand{\pderN}[3]{ \frac{\partial^{#3} #1}{\partial #2^{#3}} }

\begin{document}

\title[The solar corona as an active medium for magnetoacoustic waves]{The solar corona as an active medium for magnetoacoustic waves}

\author{D.~Y.~Kolotkov$^{1, 2}$, D.~I.~Zavershinskii$^{3, 4}$, V.~M.~Nakariakov$^{1, 5}$}

\address{$^1$ Centre for Fusion, Space and Astrophysics, Physics Department, University of Warwick, Coventry CV4 7AL, United Kingdom}
\address{$^2$ Institute of Solar-Terrestrial Physics SB RAS, Irkutsk, 664033, Russia}
\address{$^3$Department of Physics, Samara National Research University, Moscovskoe sh. 34, Samara, 443086, Russia}
\address{$^4$Department of Theoretical Physics, Lebedev Physical Institute, Novo-Sadovaya st. 221, Samara, 443011, Russia}
\address{$^5$ St. Petersburg Branch, Special Astrophysical Observatory, Russian Academy of Sciences, 196140, St. Petersburg, Russia}

\ead{d.kolotkov.1@warwick.ac.uk}
\vspace{10pt}
\begin{indented}
\item[]\today
\end{indented}

\begin{abstract}
The presence and interplay of continuous cooling and heating processes maintaining the corona of the Sun at the observed one million K temperature were recently understood to have crucial effects on the dynamics and stability of magnetoacoustic waves. These essentially compressive waves perturb the coronal thermal equilibrium, leading to the phenomenon of a wave-induced thermal misbalance. Representing an additional natural mechanism for the exchange of energy between the plasma and the wave, thermal misbalance makes the corona an active medium for magnetoacoustic waves, so that the wave can not only lose but also gain energy from the coronal heating source (similarly to burning gases, lasers and masers). We review recent achievements in this newly emerging research field, focussing on the effects that slow-mode magnetoacoustic waves experience as a back-reaction of this perturbed coronal thermal equilibrium. The new effects include enhanced frequency-dependent damping or amplification of slow waves, and effective, not associated with the coronal plasma non-uniformity, dispersion. We also discuss the possibility to probe the unknown coronal heating function by observations of slow waves and linear theory of thermal instabilities. The manifold of the new properties that slow waves acquire from a thermodynamically active nature of the solar corona indicate a clear need for accounting for the effects of combined coronal heating/cooling processes not only for traditional problems of the formation and evolution of prominences and coronal rain, but also for an adequate modelling and interpretation of magnetohydrodynamic waves.
\end{abstract}

%
\vspace{2pc}

%
%
\maketitle
\ioptwocol

\section{Introduction}
\label{sec:intro}

The coronal heating problem remains one of the major puzzles in solar physics for almost eighty years since the pioneering works \cite{1941...Alfven, 1941...Edlen}, which apparently makes it the longest-standing unsolved problem in all plasma astrophysics. Indeed, estimations show that 1\,g of the coronal plasma with typical temperature of 1\,MK radiates in the optically thin regime more than $10^{11}$\,erg\,s$^{-1}$, which would lead to the cooling of the corona in a few hours unless the radiative energy losses and parallel thermal conduction towards the chromosphere are re-supplied by some unknown yet heating mechanism. In reality, the observed lifetime of typical hot coronal plasma structures is much longer than the expected radiative cooling time (see e.g. \cite{2019ARA&A..57..157C, 2014LRSP...11....4R}, for comprehensive reviews of the properties of the corona and coronal loops), which indicates the Sun's corona has to be considered as a continuously cooling and heated medium, existing because of the everlasting competition and a delicate balance between these two processes.

The question of coronal heating is traditionally associated with the dynamics of magnetohydrodynamic (MHD) waves ubiquitously present in the corona {(see e.g. \cite{2020ARA&A..58..441N}, and also \cite{2020SSRv..216..136L, 2021SSRv..217...34W, 2021SSRv..217...73N}, for comprehensive reviews of specific coronal wave modes).}
The intrinsically filamentary nature of the coronal plasma and the associated with it generation of small spatial scales are considered as key ingredients for an effective dissipation of the wave energy to heat the plasma \cite{2015RSPTA.37340269D, 2015RSPTA.37340256K}. Despite an enormous effort in modelling and observational studies of coronal heating by MHD waves (see e.g. \cite{2020A&A...638A..89G, 2020A&A...636A..40H, 2021ApJ...908..233S}, for recent works) {and sporadic indirect estimations that in some cases the wave energy could be sufficient (e.g. \cite{2017NatSR...743147S})}, the results converge to the conclusion that realistic 3D MHD models accounting for cross-field coronal plasma inhomogeneities and observed wave amplitudes are not yet capable to reproduce the heating rate required to balance the colossal radiative losses of the corona \cite{2020SSRv..216..140V}.

\begin{figure*}
	\begin{center}
		\includegraphics[width=\linewidth]{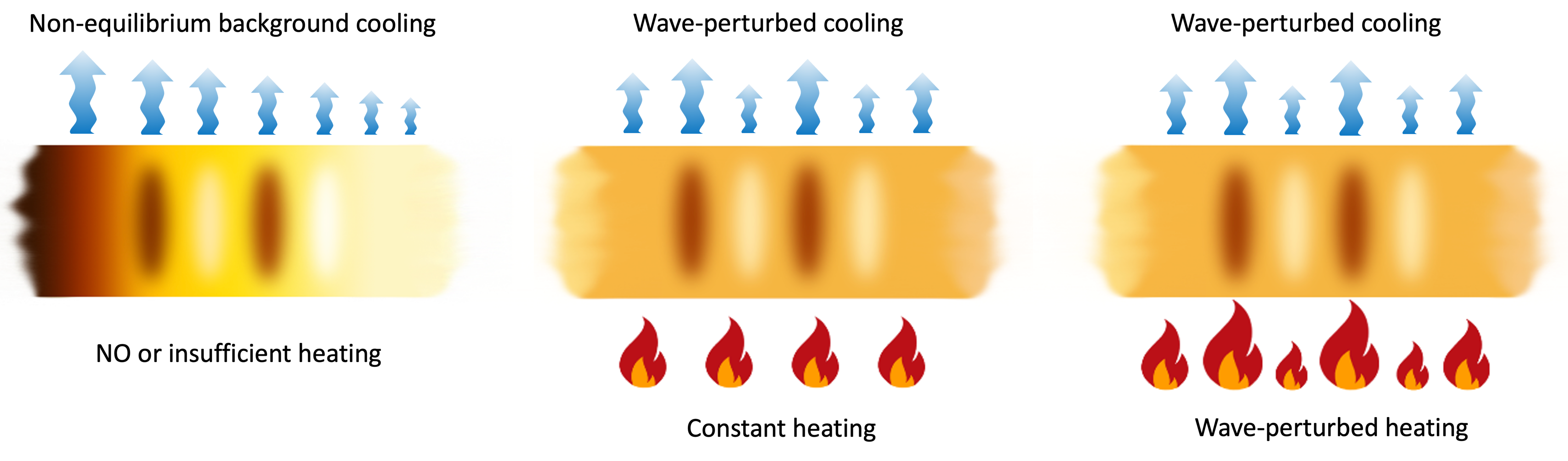}
	\end{center}
	\caption{A schematic illustration of three different scenarios for a small-amplitude compressive wave to evolve in a coronal plasma structure affected by local heating and/or cooling processes.
		Left: A wave in a coronal plasma structure continuously cooling by optically thin radiation and perhaps thermal conduction, with no local thermal equilibrium.
		Middle: A wave in a long-lived coronal plasma structure under thermal equilibrium with $\rho_0$ and $T_0$, maintained by a constant heating process ${\cal H}_0 \equiv {\cal L}_0(\rho_0,T_0)$. The wave-caused perturbations of the local plasma parameters affect the plasma cooling rate as ${\cal L}(\rho_0+\rho_1, T_0+T_1) \approx {\cal L}_0 + {\cal L}_\rho^{'}\rho_1+{\cal L}_T^{'}T_1$ and do not affect the constant heating process.
		Right: A long-lived coronal plasma structure hosting a wave, where both heating and cooling processes are functions of the local plasma parameters, ${\cal L}(\rho, T)$ and ${\cal H}(\rho, T, B)$, balancing each other in the equilibrium ${\cal H}_0(\rho_0, T_0, B_0) = {\cal L}_0(\rho_0,T_0)$. In this case, both the heating and cooling rates get modified by the wave-induced perturbations as ${\cal L}(\rho_0+\rho_1, T_0+T_1) \approx {\cal L}_0 + {\cal L}_\rho^{'}\rho_1+{\cal L}_T^{'}T_1$ and ${\cal H}(\rho_0+\rho_1, T_0+T_1, B_0 + B_1) \approx {\cal H}_0 + {\cal H}_\rho^{'}\rho_1+{\cal H}_T^{'}T_1 + {\cal H}_B^{'}B_1$, which leads to the phenomenon of wave-induced heating/cooling (thermal) misbalance.
	}
	\label{fig:sketch}
\end{figure*}

In this review, we address the link between the coronal heating and MHD waves from the wave dynamics point of view, i.e. focussing on the effects that the coronal heating and cooling processes exert on the evolution of MHD waves. In other words, here we do not consider the waves as the heating agent, but discuss the implications of combined heating/cooling effects for the wave dynamics and a practical question of seismological diagnostics of thermal properties of the corona, including its enigmatic heating function.
A vast majority of coronal wave modelling studies is carried out either in terms of ideal MHD, or with unrealistically high transport coefficients intrinsic for 3D numerical modelling.
Thus, responding to the need to go beyond ideal MHD theory for an adequate description and interpretation of the wave processes in the corona, Ref.~\cite{2008ApJ...686L.127A} considered the dynamics of fast magnetoacoustic kink waves in coronal loops undergoing rapid radiative cooling. In this approach, the hot plasma of the oscillating coronal loop was assumed to be not in a hydrostatic equilibrium, cooling down by intensive radiation which is not balanced by heating (see the schematic illustration in the left-hand panel of Fig.~\ref{fig:sketch}).
Similar studies of the effects of radiative cooling on the damping profile of kink oscillations and on the behaviour of standing slow magnetoacoustic waves were carried out in more recent works \cite{2018A&A...619A.173S, 2021FrASS...7..106S} and \cite{2013SoPh..283..413A}, respectively.
An interesting inherent feature of these models is that the coronal loop experiencing such strong radiation not compensated by heating should disappear from the observational waveband sensitive to the hot plasma emission (e.g. 94\,\AA, 171\,\AA, or 193\,\AA\ of the Atmospheric Imaging Assembly onboard Solar Dynamics Observatory, SDO/AIA) in 10--20\,min, i.e. after 2--3 cycles of oscillations. On the other hand, for example, kink oscillations of coronal loops are often observed to reside in long-lived loops for at least several clear cycles without signatures of the loop fading, or even in the decayless regime lasting for hours with more than ten oscillation cycles \cite{2013A&A...552A..57N, 2015A&A...583A.136A}.
{More details on the dissipation of kink oscillations including the effects of rapid radiative cooling are outlined in detail in the recent review \cite{2021SSRv..217...73N}.}

Accounting for a constant heating term in the coronal energy balance allowed for considering the dynamics of fast and slow magnetoacoustic waves in long-lived coronal plasma structures (see e.g. \cite{2004A&A...415..705D, 2007SoPh..246..187S, 2016ApJ...826L..20R, 2019A&A...624A..96C}). In this scenario (see the middle panel in Fig.~\ref{fig:sketch}), the role of a constant heating term reduces to maintaining the initial thermal equilibrium of the coronal plasma, neither contributing to the wave dynamics nor being affected by it. The ratio of the wave period to the characteristic timescale of the optically thin radiative cooling was shown to determine the dynamic properties of the wave in this regime.

On the other hand, taking the link between the properties of coronal loops and the local plasma parameters into account, the coronal heating rate is often modelled as a function of the local plasma density and temperature (see e.g. \cite{1978ApJ...220..643R, 1988SoPh..117...51D, 1993ApJ...415..335I, 2006A&A...460..573C}) and local value of the magnetic field strength \cite{1992PPCF...34..411H}. Moreover, the dependences of the coronal heating and radiative cooling rates on the background plasma parameters are likely to be different. Hence, perturbations of the plasma parameters caused by essentially compressive (for example, slow magnetoacoustic) waves disturb not only the mechanical equilibrium of the coronal plasma, but also violate its thermal balance through the modification of both the heating and cooling rates, thus leading to the phenomenon of a wave-induced heating/cooling (thermal) misbalance (see the right-hand panel of Fig.~\ref{fig:sketch}). Effects of the perturbed thermal equilibrium have been traditionally considered in the context of wave propagation and stability in the interstellar medium and molecular clouds (see e.g. \cite{1965ApJ...142..531F, 1977ApJ...211..400O} and more recent works \cite{2011Ap&SS.334...35M, 2017MNRAS.469.1403K}); electric discharges, lasers, and chemically-active media (e.g. \cite{1988ZhETF..94..128M}); and as mechanisms for formation and evolution of rapid condensations in the solar corona such as prominences and coronal rain (see e.g. \cite{1988SoPh..117...51D, 2011ApJ...737...27X, 2017ApJ...845...12K} and a recent review \cite{2020PPCF...62a4016A}).

In a series of more recent works, the phenomenon of local thermal misbalance was understood to strongly affect the dynamics and stability of essentially compressive slow magnetoacoustic waves in the solar corona too. Indeed, being abundantly present in the corona (see \cite{2009SSRv..149...65D, 2019ApJ...874L...1N, 2021SSRv..217...34W}, for reviews), slow waves were shown to experience a back-reaction from the perturbed thermal equilibrium, either losing or gaining energy from the coronal heating source \cite{2017ApJ...849...62N}, which makes the continuously heated and cooling corona an \emph{active medium} for magnetoacoustic waves (cf. burning gases, lasers and masers). Accounting for this intrinsic feature of the corona allowed for revealing a number of important properties of slow waves that were missing in previous models. These newly revealed properties include new mechanisms for enhanced and frequency-dependent damping of slow coronal waves \cite{2019A&A...628A.133K, 2021A&A...646A.155D, 2021SoPh..296...20P};
{modification of the phase behaviour of propagating slow waves \cite{2021SoPh..296..105P};}
a new, not associated with the plasma non-uniformity mechanism, dispersion of slow waves \cite{2019PhPl...26h2113Z, 2021arXiv210710600B}; formation of self-sustained nonlinear periodic wave structures and autosolitons (e.g. \cite{1999PhLA..254..314N, 2010PhPl...17c2107C, 2020PhRvE.101d3204Z}); and a new tool for probing the unknown coronal heating function by slow waves \cite{2020A&A...644A..33K}.

In this review, we will address some of these recent findings, focussing on the description of slow coronal waves in terms of a thin flux tube model with thermal misbalance (Sec.~\ref{sec:ttmodel}), stability of slow and entropy waves in a thermodynamically active plasma of the solar corona and its implications for diagnostics of the coronal heating function (Sec.~\ref{sec:stability}), and dispersion of slow waves, caused by thermal misbalance, and formation of quasi-periodic slow wave trains (Sec.~\ref{sec:dispersion}). A digest of the results presented and future prospects are outlined in Sec.~\ref{sec:conc}.

\section{Thin flux tube model with thermal misbalance}
\label{sec:ttmodel}

The dynamics of long-wavelength axisymmetric magnetoacoustic (MA) waves in solar coronal loops can be described in terms of a so-called thin flux tube approximation \cite{1996PhPl....3...10Z}. This approach allows one to reduce the full set of MHD equations in cylindrical coordinates to a one-dimensional form, using the second-order Taylor expansion of the variables with respect to the radial coordinate and treating the ratio of the loop radius to the characteristic wavelength as a small parameter. Thus, assuming an untwisted and non-rotating flux tube stretched along the $z$-axis, the  set of linearised equations describing the dynamics of MA waves can be written as
\begin{equation} \label{Eq_Continuity}
	\pder{\rho_1}{t} +2 \rho_0 v_{r1} + \rho_0 \pder{u_1}{z}  =0,
\end{equation}
\begin{equation} \label{Eq_Motion}
	\rho_0 \pder{u_1}{t} + \pder{P_1}{z}  =0,
\end{equation}
\begin{eqnarray} \label{Eq_Heat_Trasfer}
	C_\mathrm{V} \rho_0 \pder{T_1}{t} - \frac{k_{B} T_0}{m}  \pder{\rho_1}{t} \nonumber\\ 
	 = - \rho_0 \left(Q_{\rho}\rho_1 + Q_{T} T_1 + Q_{B} B_\mathrm{1} \right)  + \kappa_\parallel \pderN{T_1}{z}{2},
\end{eqnarray}
\begin{equation} \label{Eq_State}
	P_1  - \frac{k_{B} }{m} \left(\rho_0 T_1 +T_0 \rho_1 \right)  =0,
\end{equation}
\begin{equation} \label{Eq_Pressure_balance}
	P_1+\frac{B_0 B_1}{4 \pi} - \frac{A_0}{2 \pi} \left(\rho_0 \pder{v_{r1}}{t} + \frac{B_0}{8 \pi} \pderN{B_1}{z}{2}  \right) = 0,
\end{equation}
\begin{equation} \label{Eq_Mag_induction}
	\pder{B_1}{t} + 2 B_0 v_{r1}   =0.
\end{equation}
In Eqs.~(\ref{Eq_Continuity})--(\ref{Eq_Mag_induction}), subscripts \lq\lq 0\rq\rq\ and \lq\lq 1\rq\rq\ indicate the unperturbed value and perturbation of the corresponding variable, respectively, $\rho$ is the plasma density, $T$ is the temperature, and $P$ is the gas pressure. Also, $u$ and $B$ are the plasma velocity and magnetic field along the flux tube, $v_{r}$ is the radial derivative of the radial velocity taken at $r=0$, $A_0$ is the flux tube cross-section area. We denote ${k_B}$ for the Boltzmann constant, $C_\mathrm{V}$ for the specific heat capacity, and $m$ for the mean particle mass (see also Eq.~(\ref{eq:pars}) for the set of typical coronal plasma parameters). The set of Eqs.~(\ref{Eq_Continuity})--(\ref{Eq_Mag_induction}) is similar to that used by Ref.~\cite{2021arXiv210710600B}.

The non-adiabatic processes of coronal heating ${\cal H}(\rho, T, B)$, optically thin radiative cooling ${\cal L}(\rho, T)$, and field-aligned thermal conduction with the characteristic coefficient $\kappa_\parallel$ are accounted for on the RHS of energy Eq.~(\ref{Eq_Heat_Trasfer}). More specifically, we describe the interplay between the coronal heating and radiative cooling rates through the net heat-loss function $Q(\rho, T, B)={\cal L}(\rho, T)-{\cal H}(\rho, T, B)$, measured in W\,kg$^{-1}$ (or erg\,g$^{-1}$\,s$^{-1}$). In the equilibrium, it is equal to zero, $Q(\rho_0, T_0, B_0)=0$, providing the local thermal balance.
After a small-amplitude wave-induced perturbation, the coronal heat-loss function can be Taylor-expanded as $Q(\rho, T, B) \approx Q_{\rho}\rho_1 + Q_{T}T_1 + Q_{B}B_1$, where 
the partial derivatives $Q_{\rho} =\partial Q  / \partial \rho $, $Q_{T} =  \partial Q  / \partial T $, and $Q_{B} = \partial Q  / \partial B$ are evaluated at the initial equilibrium. The unperturbed loop is assumed isothermal along the axis, which is typical for the coronal part of solar atmospheric flux tubes.

\begin{figure}
	\begin{center}
		\includegraphics[width=\linewidth]{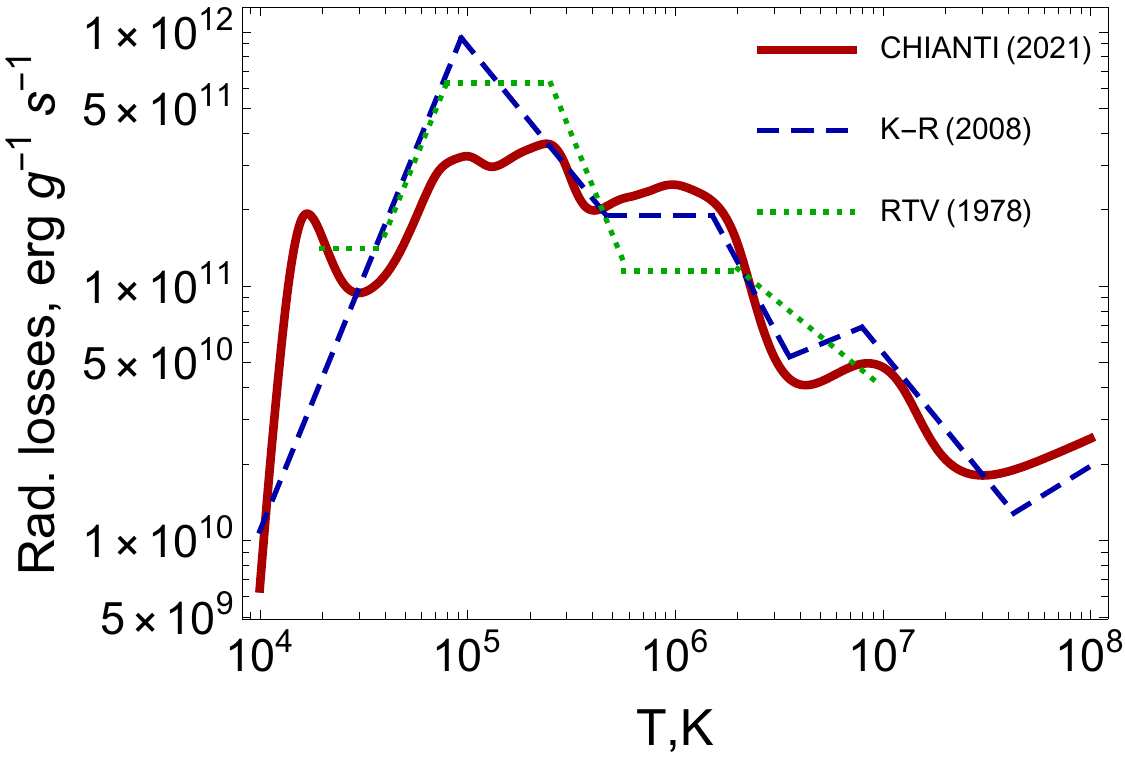}
	\end{center}
	\caption{Optically thin radiative losses per unit mass as function of the plasma temperature, obtained from the CHIANTI v.~10.1 atomic database \cite{2021ApJ...909...38D} for the coronal abundance and number density $n_0 = 10^{9}\,\mathrm{cm}^{-3}$ (the red solid line); Klimchuk--Raymond model (\cite{2008ApJ...682.1351K}, the blue dashed line); and Rosner--Tucker--Vaiana model (\cite{1978ApJ...220..643R}, the green dotted line).}
	\label{fig:radloss}
\end{figure}

It is important to stress that the heat-loss derivatives $Q_{\rho}$, $Q_{T}$, and $Q_{B}$ could be considered as effective transport coefficients of the coronal plasma, carrying the information about the coronal heating process. Indeed, using some \emph{a priori} prescribed model for optically thin radiation per unit mass, ${\cal L}(\rho, T) \propto \rho T^{\alpha}$ (for radiation per unit volume, it is $\propto \rho^2 T^{\alpha}$) and representing the unknown coronal heating function in a generic form ${\cal H}(\rho, T, B) \propto \rho^aT^bB^c$ with the power-law indices $a$, $b$, and $c$ being free dimensionless parameters, those coefficients $Q_{\rho}$, $Q_{T}$, and $Q_{B}$ become
\begin{equation}\label{eq:q_rho}
	Q_\rho = \frac{{\cal L}_0}{\rho_0}(1-a),
\end{equation}
\begin{equation}\label{eq:q_T}
	Q_T = \frac{\partial {\cal L}_0}{\partial T} - \frac{{\cal L}_0}{T_0}b,
\end{equation}
\begin{equation}\label{eq:q_B}
	Q_B = - \frac{{\cal L}_0}{B_0}c.
\end{equation}
As one can see from Eqs.~(\ref{eq:q_rho})--(\ref{eq:q_B}), the coefficients $Q_{\rho}$, $Q_{T}$, and $Q_{B}$ and hence their effect on the wave dynamics depend on the unknown heating parameters $a$, $b$, and $c$ and, on the other hand, are highly sensitive to the value and local gradients of the unperturbed radiative cooling function ${\cal L}_0$. In Fig.~\ref{fig:radloss}, we illustrate the dependence of the coronal radiative losses ${\cal L}_0$ on the plasma temperature, as it is predicted by the Rosner--Tucker--Vaiana \cite{1978ApJ...220..643R}, Klimchuk--Raymond \cite{2008ApJ...682.1351K}, and CHIANTI \cite{2021ApJ...909...38D} models.
Despite the similarity in a general tendency of the function ${\cal L}_0$ to vary with temperature, the considered models have clearly different local gradients of ${\cal L}_0$.
In this work, we use the most recent and therefore accurate version 10.1 of the CHIANTI model for numerical examples.

The original set of Eqs.~(\ref{Eq_Continuity})--(\ref{Eq_Mag_induction}) can be reduced to a single equation describing the dynamics of linear MA and entropy waves in a plasma with thermal misbalance (TM) and parallel thermal conduction (TC), under the thin flux tube approximation,
\begin{equation} \label{waveEq_general}
	\left[\pder{}{t} {\cal D}_\mathrm{ideal} + \frac{ Q_{T}}{C_\mathrm{V}}{\cal D}_\mathrm{TM} - \frac{\kappa_\parallel}{\rho_0 C_\mathrm{V}} \pderN{}{z}{2} {\cal D}_\mathrm{TC}\right] \rho_1  =0,
\end{equation}
with the following differential operators,
\begin{eqnarray} \label{Operators_general}
	{\cal D}_\mathrm{ideal}=\left(c_\mathrm{A}^2+c_\mathrm{s}^2\right)\pderN{}{t}{2}-c_\mathrm{s}^2 c_\mathrm{A}^2\pderN{}{z}{2}\nonumber\\
	 +\frac{A_0}{4\pi}\left(\pderN{}{t}{2}-c_\mathrm{s}^2\pderN{}{z}{2}\right)\left(\pderN{}{t}{2}-c_\mathrm{A}^2\pderN{}{z}{2}\right),\nonumber\\
	{\cal D}_\mathrm{TM}=\left(c_\mathrm{A}^2+ \frac{Q_{T} T_0 - Q_{\rho}\rho_0 - Q_{B}B_0  }{\gamma Q_{T} T_0 } c_\mathrm{s}^2  \right)\pderN{}{t}{2} \nonumber\\
	 - \frac{Q_{T} T_0 - Q_{\rho}\rho_0 }{\gamma Q_{T} T_0 } c_\mathrm{s}^2 c_\mathrm{A}^2 \pderN{}{z}{2}\nonumber\\
	+\frac{A_0}{4\pi}\left(\pderN{}{t}{2}- \frac{Q_{T}T_0 - Q_{\rho}\rho_0  }{ \gamma Q_{T}T_0 } c_\mathrm{s}^2 \pderN{}{z}{2}\right)\left(\pderN{}{t}{2}-c_\mathrm{A}^2\pderN{}{z}{2}\right),\nonumber\\ 
	{\cal D}_\mathrm{TC}=\left(c_\mathrm{A}^2+\frac{c_\mathrm{s}^2}{\gamma}\right)\pderN{}{t}{2}-\frac{c_\mathrm{s}^2 c_\mathrm{A}^2}{\gamma}\pderN{}{z}{2}\nonumber\\
	 +\frac{A_0}{4\pi}\left(\pderN{}{t}{2}-\frac{c_\mathrm{s}^2}{\gamma}\pderN{}{z}{2}\right)\left(\pderN{}{t}{2}-c_\mathrm{A}^2\pderN{}{z}{2}\right),\nonumber 
\end{eqnarray}
where $c_\mathrm{A} = B_0/ \sqrt{4 \pi \rho_0}$ and $c_{\mathrm{s}} =\sqrt{\gamma {k}_{B}  T_0 /m} $ are the standard Alfv\'en and sound (with the adiabatic index $\gamma=5/3$) speeds in the plasma.

Equation~(\ref{waveEq_general}) is seen to be of the fifth-order in time, so that it describes the dynamics of two fast MA modes, two slow MA modes, and one entropy mode. However, as any change in the external (outside the flux tube) pressure is neglected on the RHS of Eq.~(\ref{Eq_Pressure_balance}) (see e.g. \cite{1983SoPh...88..179E}), we focus on the dynamics of essentially compressive slow MA and entropy waves, which propagate/evolve inside the loop, always in a trapped regime. For the use of this model for an adequate description of fast MA waves, the interaction of the flux tube with the external medium should be taken into account (see e.g. \cite{2014ApJ...781...92V}).

In the case of  weak  thermal conduction $\kappa_\parallel\to 0$ and $Q_B\to 0$ (no dependence of the coronal heating rate on the local magnetic field strength), Eq.~(\ref{waveEq_general}) coincides with that derived in Ref.~\cite{2021arXiv210710600B}. Likewise, neglecting the effects of finite loop width $(A_0 k^2 \to 0)$, Eq.~(\ref{waveEq_general}) reduces to the model of slow MA waves affected by thermal misbalance and thermal conduction, considered by Ref.~\cite{2021A&A...646A.155D}. If one additionally assumes the loop's magnetic field is infinitely strong with $c_\mathrm{A}\gg c_\mathrm{s}$, Eq.~(\ref{waveEq_general}) yields a so-called infinite field approximation, in which slow MA waves do not perturb the magnetic field and its role effectively reduces to determining the propagation direction and 1D nature of the wave (see e.g. \cite{2019A&A...628A.133K, 2019PhPl...26h2113Z, 2021SoPh..296...96Z}).



\section{Stability of acoustic and thermal modes and coronal heating function}
\label{sec:stability}

In a highly magnetised plasma with $c_\mathrm{A}\gg c_\mathrm{s}$ typical for the solar corona, the dynamics of slow MA waves was shown to be insensitive to the dependence of the unknown coronal heating function on the local magnetic field strength (see Fig.~2 in Ref.~\cite{2021A&A...646A.155D}). This reduces the parametrisation of the coronal heating function through the local plasma parameters to ${\cal H}(\rho,T) \propto \rho^a T^b$, with two unknowns $a$ and $b$. In this low-$\beta$ regime, Ref.~\cite{2020A&A...644A..33K} obtained the following conditions for thermal (entropy) and slow MA modes described by Eq.~(\ref{waveEq_general}) to remain stable,
\begin{eqnarray}
	&a - b + \frac{T_0}{{\cal L}_0}\frac{\partial {\cal L}_0}{\partial T} + \frac{C_\mathrm{V}}{\tau_\mathrm{c}(k)}\frac{T_0}{{\cal L}_0} - 1 > 0, &\label{eq:instab_cond_th_ab}\\
	&-\frac{a}{\gamma-1} - b + \frac{T_0}{{\cal L}_0}\frac{\partial {\cal L}_0}{\partial T} + \frac{C_\mathrm{V}}{\tau_\mathrm{c}(k)}\frac{T_0}{{\cal L}_0} +\frac{1}{\gamma-1} > 0,&\label{eq:instab_cond_ac_ab}
\end{eqnarray}
respectively. Here, $\tau_\mathrm{c}(k)={\rho_0 C_\mathrm{V}k^{-2}}/{\kappa_\parallel}$ is the wavelength-dependent characteristic time of the parallel thermal conduction, and the radiative loss function ${\cal L}(\rho, T)$ is illustrated in Fig.~\ref{fig:radloss}. Equations~(\ref{eq:instab_cond_th_ab})--(\ref{eq:instab_cond_ac_ab}) demonstrate that the term associated with the field-aligned thermal conduction is always positive so it tends to stabilise the perturbation and restore the initial equilibrium, while the terms associated with thermal misbalance (namely, the heating power-law indices $a$ and $b$, and the local gradient of the radiative loss function $\partial {\cal L}_0/\partial T$) as well as their combinations could be both positive and negative. The latter implies that for different coronal plasma conditions and different coronal heating models, the phenomenon of thermal misbalance may lead to a number of different scenarios for the initial compressive perturbation to evolve in the solar corona, with the acoustic and thermal modes being stable (decaying) and/or unstable.

\begin{figure}
	\begin{center}
		\includegraphics[width=\linewidth]{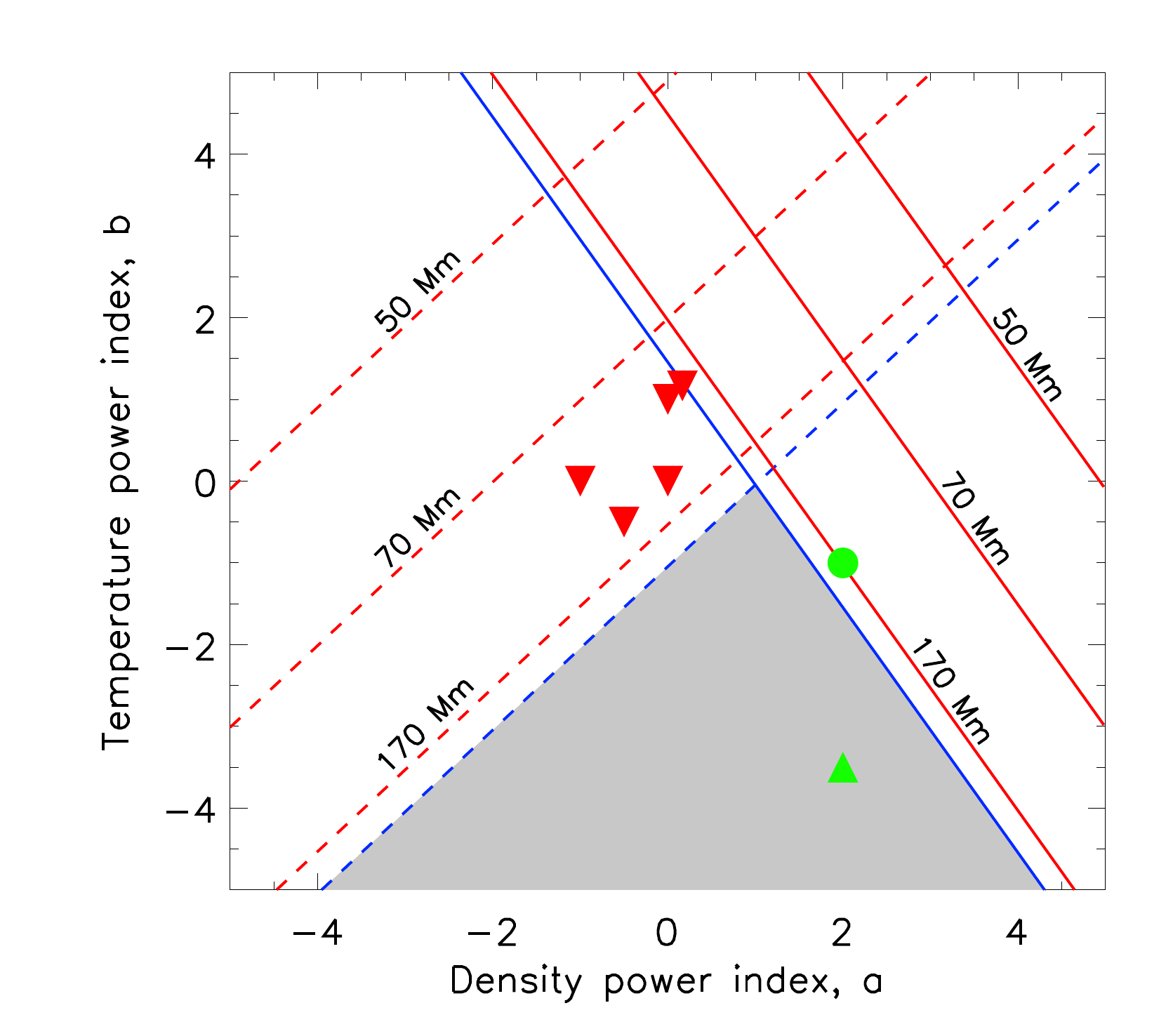}
	\end{center}
	\caption{The analysis of stability described by Eqs.~(\ref{eq:instab_cond_th_ab})--(\ref{eq:instab_cond_ac_ab}) of thermal (entropy) and acoustic modes in the coronal plasma (\ref{eq:pars}) with $\beta\to0$, affected by thermal conduction and thermal misbalance with the unknown coronal heating function ${\cal H}\propto\rho^aT^b$ (the dependence of the heating rate on the magnetic field has no effect on the wave dynamics in the considered low-$\beta$ regime). For the heating models from the grey-shaded region, both thermal and acoustic modes are stable independently of the wavelength. Outside this region, the corona is stable to the perturbations shorter than the thermal and acoustic Field's lengths (\ref{eq:lambda_thermal})--(\ref{eq:lambda_acoustic}) shown by dashed and solid lines, respectively. The red triangles show the RTV heating models \cite{1978ApJ...220..643R}. The green symbols show the heating models used for illustrations in Sec.~\ref{sec:dispersion}.
	}
	\label{fig:stability_ab}
\end{figure}

The regions of the heating power-indices $a$ and $b$, corresponding to stable/unstable behaviour of the acoustic and thermal modes, prescribed by Eqs.~(\ref{eq:instab_cond_th_ab})--(\ref{eq:instab_cond_ac_ab}), are demonstrated in Fig.~\ref{fig:stability_ab} for the following combination of the plasma parameters common for many typical wave-hosting structures in the solar corona,

\setlength{\arraycolsep}{0pt}
\begin{equation}\label{eq:pars}
	\left\{\begin{array}{l} 
	\mathrm{Temperature},~T_0= 1\,\mathrm{MK},\\[0.1cm]
	\mathrm{Number~density},~n_0 = 10^{9}\,\mathrm{cm}^{-3},\\[0.1cm]
	\mathrm{Magnetic~field},~B_0 = 4\,\mathrm{G}~(\beta\approx 0.2),~40\,\mathrm{G}~(\beta\to 0), \\[0.1cm]
	\mathrm{Parallel~thermal}\\[-0.17cm]
	\mathrm{conductivity},~\kappa_\parallel=10^{-11}T_0^{5/2}\, \mathrm{W\,m}^{-1}\,\mathrm{K}^{-1},\\[0.1cm]
	\mathrm{Mean~particle~mass},~m=0.6\times1.67\times10^{-27}\,\mathrm{kg},\\[0.1cm]
	\mathrm{Adiabatic~index},~\gamma =5/3,\\[0.1cm]
	\mathrm{Specific~heat~capacity},~C_\mathrm{V}=\displaystyle\frac{k_\mathrm{B}}{(\gamma-1)m}\,\mathrm{J}\,\mathrm{K}^{-1}\,\mathrm{kg}^{-1},\\[0.1cm]
	\mathrm{Wavelength},~\lambda=50\,\mathrm{Mm},
	\end{array}\right.
\end{equation}
which gives the standard sound speed $c_\mathrm{s} \approx 152$\,km\,s$^{-1}$ and the corresponding acoustic period $P_\mathrm{A}=\lambda/c_\mathrm{s}\approx 5.5$\,min, used as characteristic properties of slow waves in the ideal plasma for normalisation.

The grey-shaded region in Fig.~\ref{fig:stability_ab} shows values of the heating power-indices $a$ and $b$, for which both the effect of thermal misbalance and parallel thermal conduction lead to the decay of thermal and acoustic modes. As such, the stability of the coronal plasma is wavelength-independent in this domain. In contrast, for the values of $a$ and $b$ outside the grey-shaded region in Fig.~\ref{fig:stability_ab}, thermal misbalance leads to an effective gain of energy by the perturbation from the coronal heating source, counteracting the damping by thermal conduction or even causing the amplification of thermal and acoustic modes. In this region, the stability of the coronal plasma depends on the wavelength of the perturbation. Thus, for some guessed heating model with $a$ and $b$, there is a critical wavelength $\lambda_\mathrm{F}$ below which thermal conduction is strong enough to suppress the instability of these harmonics, caused by thermal misbalance. For longer-wavelength perturbations, thermal conduction is too weak to restrain the effect of thermal misbalance and thus the coronal plasma gets unstable to these perturbations. In particular, for typical coronal parameters (\ref{eq:pars}), the RTV heating models \cite{1978ApJ...220..643R} are seen to be stable to the perturbations shorter than $\sim$100\,Mm.

From Eq.~(\ref{eq:instab_cond_th_ab}), the critical wavelength at which thermal conduction becomes insufficient to sustain the instability of thermal (entropy) mode is
\begin{equation}\label{eq:lambda_thermal}
	\lambda_\mathrm{F}^\mathrm{thermal}=2\pi\sqrt{\frac{\kappa_\parallel T_0}{\rho_0{\cal L}_0\left[1-a+b-\displaystyle\frac{T_0}{{\cal L}_0}\frac{\partial {\cal L}_0}{\partial T}\right]}},
\end{equation}
which could be understood as a \emph{thermal Field's length} generalised for a non-constant coronal heating rate (cf. Eq.~(1) and its discussion in \cite{2020PPCF...62a4016A}). Likewise, using Eq.~(\ref{eq:instab_cond_ac_ab}), one can obtain the wavelength at which the acoustic mode grows,
\begin{equation}\label{eq:lambda_acoustic}
	\lambda_\mathrm{F}^\mathrm{acoustic}=2\pi\sqrt{\frac{\kappa_\parallel T_0}{\rho_0{\cal L}_0\left[\displaystyle\frac{a-1}{\gamma-1}+b-\displaystyle\frac{T_0}{{\cal L}_0}\frac{\partial {\cal L}_0}{\partial T}\right]}},
\end{equation}
which can be referred to as an \emph{acoustic Field's length}.

\section{Non-waveguide dispersion of slow magnetoacoustic waves}
\label{sec:dispersion}

Searching for the solution of Eq.~(\ref{waveEq_general}) in a harmonic form $\propto{e}^{{i}(kz - \omega t)}$ with the cyclic frequency $\omega$ and the wavenumber $k$ yields the following dispersion relation,
\begin{eqnarray} \label{Disp_General}
	A_0 k^2 \left[\omega^5+\Delta_4(k)\omega^4\right] +\Delta_3(k)\omega^3+\\
	 +\Delta_2(k)\omega^2+\Delta_1(k)\omega+\Delta_0(k)=0,\nonumber
\end{eqnarray}
where the coefficients are
\begin{eqnarray} \label{coef_Disp_General}
	\Delta_0(k)= \frac{{i}  k ^4 c_\mathrm{A}^2 c_\mathrm{s}^2}{\gamma} \left({A_0}k^2+{4 \pi}\right) \left( \frac{k^2\kappa_\parallel}{C_\mathrm{V} \rho_0} + \frac{Q_{T} T_0 - Q_{\rho}\rho_0}{C_\mathrm{V} T_0} \right),  \nonumber\\
	\Delta_1(k)= \left({A_0}k^2+{4 \pi}\right) c_\mathrm{A}^2 c_\mathrm{s}^2 k^4 , \nonumber\\
	\Delta_2(k)=-{i} k^2 \Bigg\{ \left({A_0}k^2+{4 \pi}\right) \bigg[  \frac{k^2\kappa_\parallel}{C_\mathrm{V} \rho_0}  \left(c_\mathrm{A}^2+\frac{c_\mathrm{s}^2}{\gamma}\right)
	\nonumber\\ 
	+ \frac{Q_T}{C_\mathrm{V}} \left(c_\mathrm{A}^2+ \frac{Q_{T} T_0 - Q_{\rho}\rho_0  }{\gamma Q_{T} T_0 } c_\mathrm{s}^2  \right) \bigg] - {4 \pi} \frac{c_\mathrm{s}^2 B_0 Q_B}{\gamma C_\mathrm{V} T_0}
	\Bigg\}, \nonumber\\
	\Delta_3(k)= -k^2 \left(A_0 k^2+4 \pi\right)  \left(c_\mathrm{A}^2+c_\mathrm{s}^2\right ), \nonumber\\
	\Delta_4(k)={i} \left(\frac{Q_T}{C_\mathrm{V}}+\frac{k^2\kappa_\parallel}{C_\mathrm{V} \rho_0}  \right). \nonumber
\end{eqnarray}
In terms of the considered model, the main sources of the wave dispersion described by Eq.~(\ref{Disp_General}) are the finite cross-section of the waveguiding flux tube ($A_0 k^2$, i.e. the geometrical dispersion), thermal conduction ($\kappa_\parallel$), and the new dispersion caused by the effect of coronal heating/cooling misbalance ($Q_\rho$, $Q_T$, and $Q_B$). For example, neglecting the effects of waveguiding dispersion ($A_0 k^2 \to 0$), Eq.~(\ref{Disp_General}) can be resolved for slow MA waves as
\begin{equation}\label{eq:omega_r}
	\omega_\mathrm{R} \approx c_\mathrm{T}k,
\end{equation}
\begin{equation}\label{eq:omega_im}
	\omega_\mathrm{I} \approx-\frac{1}{2}\left(\frac{2}{2+\gamma\beta}\right)\left[\frac{\gamma-1}{\gamma}\frac{1}{\tau_c(k)}+\frac{1}{\tau_2}-\frac{1}{\tau_1}\right],
\end{equation}
where $\omega_\mathrm{R}$ and $\omega_\mathrm{I}$ are real and imaginary parts of the complex cyclic frequency $\omega=\omega_\mathrm{R}+i\omega_\mathrm{I}$, $c_\mathrm{T}=c_\mathrm{s}c_\mathrm{A}/\sqrt{c_\mathrm{s}^2+c_\mathrm{A}^2}$ is the standard tube speed, $\beta=(2/\gamma)(c_\mathrm{s}^2/c_\mathrm{A}^2)$, $\tau_\mathrm{c}(k)={\rho_0 C_\mathrm{V}k^{-2}}/{\kappa_\parallel}$ is the wavelength-dependent characteristic time of the field-aligned thermal conduction, and
\begin{equation}\label{eq:tau_12}
	\tau_1 = \frac{\gamma C_\mathrm{V}}{Q_{T} - Q_{\rho}\rho_0 / T_0 },~~\tau_2 = \frac{C_\mathrm{V}}{Q_{T} - \beta Q_{B} B_0 / 2 T_0 }
\end{equation}
are the characteristic timescales of thermal misbalance, obtained through the temperature-derivatives of the coronal heat-loss function, taken at constant gas and magnetic pressures, respectively. For typical coronal plasma conditions, the timescales $\tau_{1,2}$ were shown to be about several minutes, thus coinciding with periods and damping times of slow MA waves observed in the solar corona \cite{2020A&A...644A..33K}. Solution (\ref{eq:omega_r})--(\ref{eq:omega_im}) was obtained by Ref.~\cite{2021A&A...646A.155D} in the limit of weak non-adiabaticity, i.e. treating all non-adiabatic processes slow in comparison with the wave period, providing $\omega\tau_i\gg 1$ and $\omega_\mathrm{R}\gg\omega_\mathrm{I}$. Similar solutions in the regime of strong non-adiabaticity with $\omega\tau_i\ll 1$ could also be found in Ref.~\cite{2021A&A...646A.155D}.


\begin{figure*}
	\begin{center}
		\includegraphics[width=\linewidth]{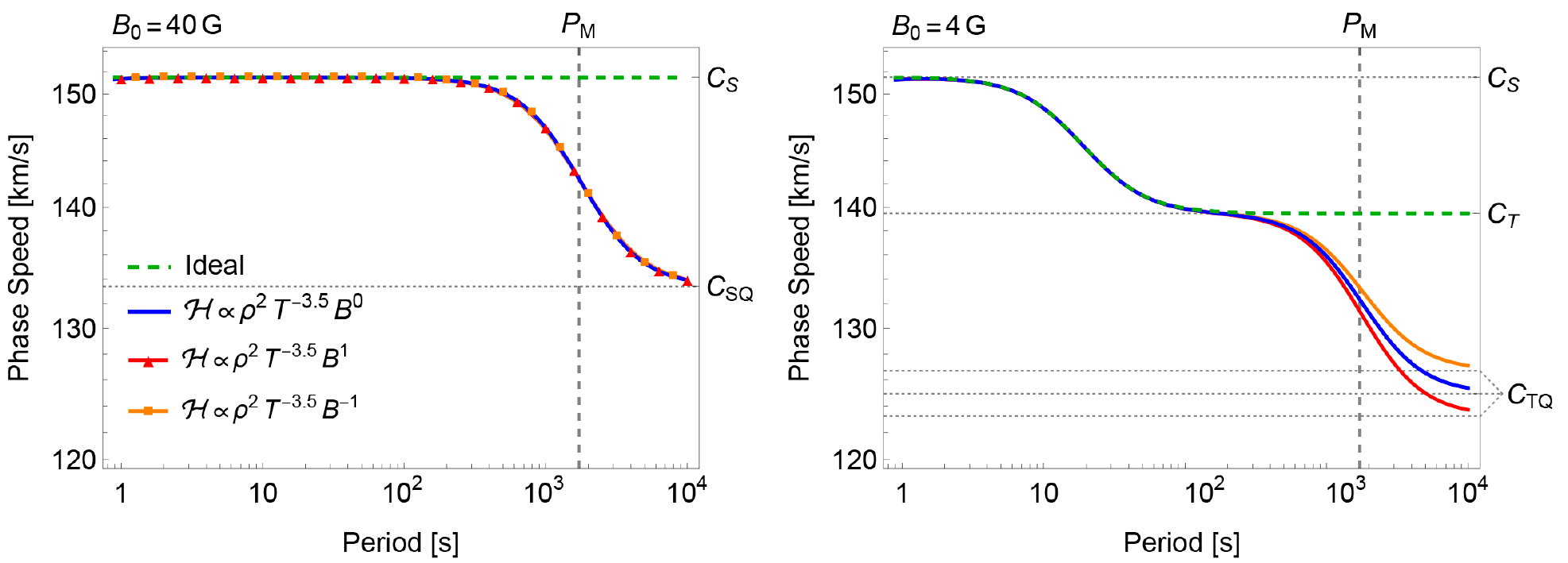}
	\end{center}
	\caption{The dependence of the phase speed of a slow MA wave on the wave period in a low-$\beta$ (left) and finite-$\beta$ (right) coronal plasma tube with parameters (\ref{eq:pars}), with and without thermal misbalance. The colour scheme is the same for both panels. The dispersion curves with thermal misbalance are calculated for three different heating models, in which the dependence on density and temperature is taken from the stability analysis (Fig.~\ref{fig:stability_ab}), and the dependences on the magnetic field are chosen arbitrarily for illustration. The long-period phase speed limits $c_\mathrm{sQ}$ and $c_\mathrm{TQ}$ are given by Eqs.~(\ref{C_SQ}) and (\ref{C_TQ}). The wave period $P_\mathrm{M}$ at which the misbalance-caused dispersion is most effective is given by Eq.~(\ref{P_m}). See also Ref.~\cite{2021arXiv210710600B}, for the case with ${\cal H}\propto\rho^{0.5}T^{-3.5}B^{0}$.
	}
	\label{phase_speed_plot}
\end{figure*}


In reality, slow MA waves are usually observed in the solar corona to damp rapidly, within a few oscillation cycles \cite{2009SSRv..149...65D, 2019ApJ...874L...1N, 2021SSRv..217...34W}, which is equivalent to $\omega_\mathrm{R}\sim\omega_\mathrm{I}$. Hence, strictly speaking, neither weak nor strong limits of non-adiabaticity are applicable. For $\omega_\mathrm{R}\sim\omega_\mathrm{I}$, Refs.~\cite{2019PhPl...26h2113Z, 2021arXiv210710600B} demonstrated that the phenomenon of thermal misbalance leads to an effective dispersion of slow MA waves, manifested through the dependence of the wave propagation speed on the wave frequency. Thus, in Fig.~\ref{phase_speed_plot}, we illustrate the dependence of the slow MA wave phase speed $V_\mathrm{ph}=\omega_\mathrm{R}/k$ on the wave period $P=2\pi/\omega_\mathrm{R}$, obtained from the full solution of dispersion relation (\ref{Disp_General}). More specifically, the dispersion curves shown in Fig.~\ref{phase_speed_plot} are calculated for $\kappa_\parallel\to 0$, which allows for a direct comparison of the dispersive effects that slow MA waves experience from the coronal plasma waveguide and the process of wave-induced thermal misbalance. We adopt the set of coronal plasma parameters given by Eq.~(\ref{eq:pars}), the cooling rate obtained from the CHIANTI database (Fig.~\ref{fig:radloss}), and the heating rate in the form of a power-law function ${\cal H} \propto \rho^a T^b B^c$. In order to satisfy stability conditions (\ref{eq:instab_cond_th_ab})--(\ref{eq:instab_cond_ac_ab}), we take the heating model with $a=2$ and $b=-3.5$ (see Fig.~\ref{fig:stability_ab}), and consider three different dependences on the magnetic field, with $c = 1$, $0$, and $-1$.


The left-hand panel of Fig.~\ref{phase_speed_plot} shows the dependence of the phase speed $V_\mathrm{ph}$ on the wave period $P$, obtained from Eq.~(\ref{Disp_General}) in the infinite magnetic field approximation of slow MA waves, with $c_\mathrm{A}\gg c_\mathrm{s}$. In this regime, slow MA waves propagate strictly along the waveguide axis without perturbing its boundaries. In other words, the tube speed $c_\mathrm{T}$, which is the long-period limit of the phase speed in the ideal plasma (without thermal misbalance), tends to the sound speed $c_\mathrm{s}$, making the effect of the geometrical dispersion on the dynamics of slow waves negligible. Thus, the dominant mechanism of slow-wave dispersion in this regime is thermal misbalance, due to which the phase speed varies from a short-period value $c_\mathrm{s}$ to the long-period value $c_\mathrm{sQ}$, prescribed by the properties of the heating/cooling processes, plasma temperature, and density as
\begin{equation}\label{C_SQ}
	c_\mathrm{sQ}^2=\left(1-\frac{\rho_0}{T_0}\frac{Q_\rho}{Q_T}\right) \frac{k_{B} T_0}{ m}.
\end{equation}
It is worth mentioning that, in general,  $c_\mathrm{sQ}$ may be either greater or lower than  $c_\mathrm{s}$. Also, the dependence of the heating rate on the magnetic field strength has no effect on the slow-wave phase speed in this low-$\beta$ regime, which is consistent with Ref.~\cite{2021A&A...646A.155D}.

The dependence of the phase speed on the slow-wave period in the regime of finite plasma-$\beta$ is shown in the right-hand panel of Fig.~\ref{phase_speed_plot}. In this case, the geometrical dispersion of slow waves appears. In the ideal plasma, it leads to the variation of the phase speed from $c_\mathrm{s}$ at short periods to $c_\mathrm{T}$ at long periods. However, for the considered value of $\beta\approx0.2$ (see Eq.~(\ref{eq:pars})), $c_\mathrm{T}$ differs from $c_\mathrm{s}$ by several percent only, which is difficult to detect in observations. Hence, the waveguiding dispersion is usually thought to be weakly important for slow MA waves in the solar corona. However, accounting for the effect of thermal misbalance leads to stronger departure of the slow-wave phase speed from the standard $c_\mathrm{s}$ towards a new value $c_\mathrm{TQ}$ at longer periods, determined by the properties of heating/cooling processes and equilibrium plasma parameters as
\begin{equation}\label{C_TQ}
	c_\mathrm{TQ}^2 = { \frac{ c_\mathrm{sQ}^2 c_\mathrm{A}^2}{ c_\mathrm{sQ}^2  + c_\mathrm{A}^2 - Q_{B} B_0 k_{B} /  {Q_{T} m} } }. 
\end{equation}
For example, Ref.~\cite{2021arXiv210710600B} demonstrated that the relative error in the seismological estimation of the coronal magnetic field by slow waves (see e.g. \cite{2007ApJ...656..598W, 2016NatPh..12..179J}), could reach up to 40\% if one neglects the effect of misbalance and uses the standard tube speed $c_\mathrm{T}$ instead of $c_\mathrm{TQ}$. In contrast to $c_\mathrm{sQ}$ (\ref{C_SQ}), the value of $c_\mathrm{TQ}$ (\ref{C_TQ}) is shown to be sensitive to the dependence of the coronal heating function on the magnetic field strength through the heat-loss derivative $Q_B$ (\ref{eq:q_B}). 

\begin{figure}
	\begin{center}
		\includegraphics[width=\linewidth]{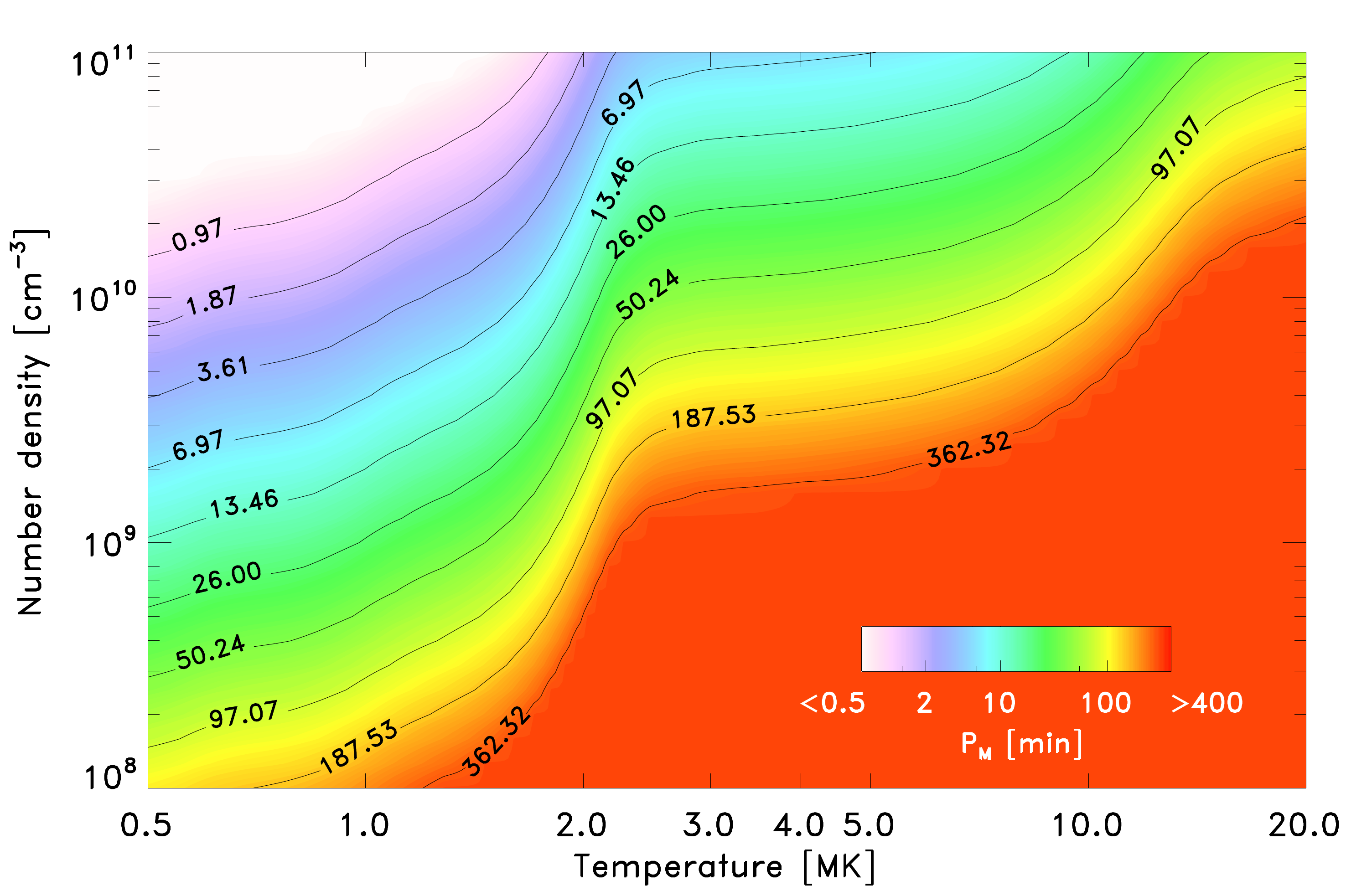}
	\end{center}
	\caption{The wave period $P_\mathrm{M}$ (\ref{P_m}) of most effective dispersion of slow MA waves, caused by thermal misbalance (see Fig.~\ref{phase_speed_plot}).
	}
	\label{fig:permisbal}
\end{figure}

The characteristic wave period at which the effect of dispersion caused by thermal misbalance is maximum was determined by Ref.~\cite{2019PhPl...26h2113Z} in the infinite field approximation ($\beta \rightarrow 0$ and $A_0 k^2 \rightarrow 0$) through the maximum gradient of the phase speed, as
\begin{eqnarray} \label{P_m}
	P_\mathrm{M} = 2\pi \sqrt{\tau_{1}\tau_{2}} =  2\pi \sqrt{\frac{\gamma C_\mathrm{V}^2}{ Q_{T} \left({Q_{T} - Q_{\rho}\rho_0 / T_0 }\right)}},
\end{eqnarray}
where $\tau_{1,2}$ are the timescales of thermal misbalance, given by Eq.~(\ref{eq:tau_12}). Moreover, the dependence of the heating rate on the magnetic field strength is seen to become important in the right-hand panel of Fig.~\ref{phase_speed_plot} at wave periods comparable to or greater than $P_\mathrm{M}$. Typical values of the wave period $P_\mathrm{M}$ are shown in Fig.~\ref{fig:permisbal}, for broad intervals of coronal plasma densities and temperatures, and $\beta\to0$. In particular, in the quiescent corona and coronal holes with typical densities $\approx$(0.5--1)$\times 10^9$\,cm$^{-3}$ and temperatures around 1\,MK, the wave period $P_\mathrm{M}$ is seen to be from 20\,min to 70\,min, which coincides with observations of very long-period propagating compressive waves in coronal plumes and inter-plume regions \cite{2001A&A...377..691B, 2001A&A...380L..39B}.

\begin{figure*}
	\begin{center}
		\includegraphics[width=\linewidth]{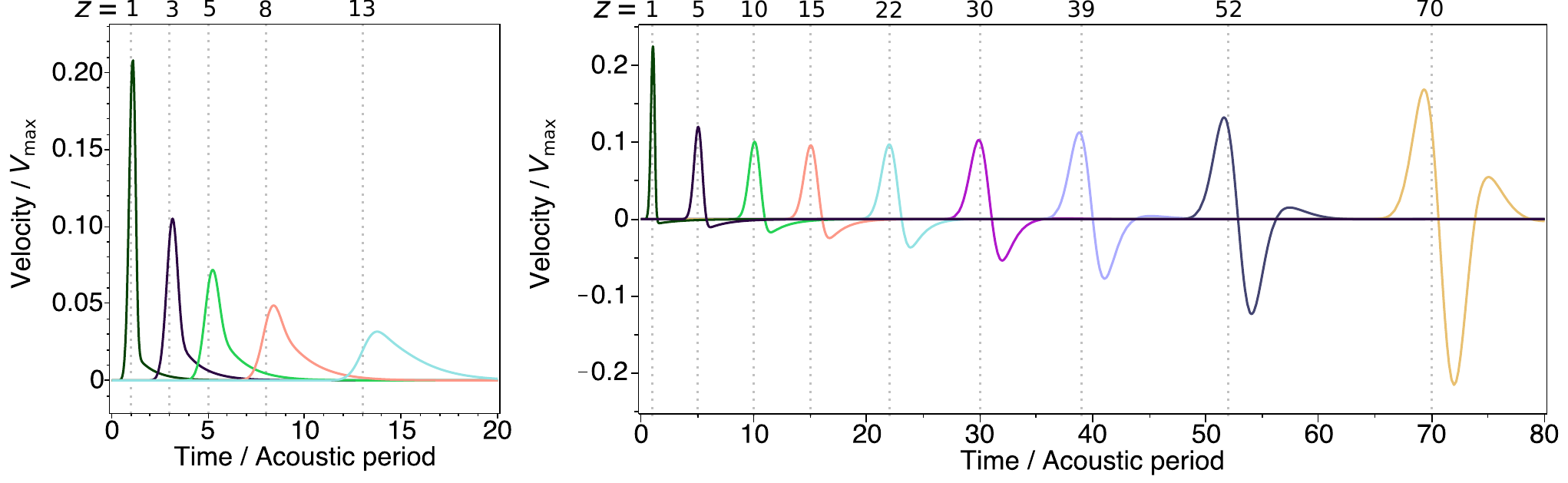}
	\end{center}
	\caption{Evolution of a broadband velocity pulse, obtained from a numerical solution of Eq.~(\ref{waveEq_general}) with $A_0 k^2 \rightarrow 0$ and $c_\mathrm{A}\gg c_\mathrm{s}$, plasma parameters (\ref{eq:pars}) with $\beta \to 0$, and heating models ${\cal H}\propto \rho^2T^{-3.5}$ (left) and ${\cal H}\propto \rho^2T^{-1}$ (right). Different colours show the same signal measured at different distances $z$ from the site of initial perturbation. The vertical dotted lines show the expected locations of the pulse in the ideal plasma, i.e. in the absence of the slow wave dispersion caused by thermal misbalance. The time and $z$-coordinate are normalised to the acoustic period $P_\mathrm{A}\approx 5.5$\,min and wavelength $\lambda=50$\,Mm, determined for plasma parameters (\ref{eq:pars}) in the ideal plasma. The signal's amplitude is normalised to the amplitude of the initial perturbation, $V_\mathrm{max}$.
	}
	\label{fig:examples}
\end{figure*}

The evolution of an essentially broadband velocity pulse of a slow-mode MA nature, affected by the above-described dispersion effects, is shown in Fig.~\ref{fig:examples}, obtained from a numerical solution of Eq.~(\ref{waveEq_general}) with $A_0 k^2 \rightarrow 0$ and $c_\mathrm{A}\gg c_\mathrm{s}$ and for plasma parameters (\ref{eq:pars}). Namely, the left-hand panel shows the regime with the heating model ${\cal H}\propto \rho^2T^{-3.5}$, for which both processes of thermal conduction and heating/cooling misbalance lead to the damping of all slow-mode (and entropy) harmonics constituting the initial perturbation (see Fig.~\ref{fig:stability_ab}). In this case, the initially localised pulse is seen to rapidly decrease in the amplitude and disperse (broaden) in time and space, due to faster propagation and more effective damping of shorter-period slow MA harmonics. In the right-hand panel of Fig.~\ref{fig:examples}, we demonstrate the evolution of the same initially localised velocity pulse, using the heating model ${\cal H}\propto \rho^2T^{-1}$, for which all entropy-mode harmonics decay, while for slow MA harmonics there is a critical wavelength $\lambda_\mathrm{F} \approx 170$\,Mm determined by Eq.~(\ref{eq:lambda_acoustic}), above which slow waves get overstable through the effective gain of energy from the coronal heating source (see Fig.~\ref{fig:stability_ab}). Thus, at the initial stage of the pulse evolution, its amplitude is seen to decrease by conductive damping of the harmonics shorter than $\lambda_\mathrm{F} \approx 170$\,Mm, followed by the amplification of longer-wavelength harmonics and formation of long-period quasi-periodic patterns by the effect of thermal misbalance. The dominant period of these slow-propagating quasi-periodic wave trains is seen to be about $P_\mathrm{M}$ (\ref{fig:permisbal}), which is approximately 26\,min for the combination of plasma parameters (\ref{eq:pars}). More examples with $\kappa_\parallel\to 0$ can be found in Ref.~\cite{2021SoPh..296...96Z}.

\section{Summary and prospects}
\label{sec:conc}

We discussed the phenomenon of a wave-induced heating/cooling (thermal) misbalance, which occurs in a continuously heated and cooling plasma of the solar corona due to the violation of a local thermal balance by compressive magnetoacoustic waves. Acting as an additional natural mechanism for the exchange of energy between the plasma and the wave, the phenomenon of thermal misbalance makes the corona an active medium for magnetoacoustic waves. The presented results and physical effects that the wave experiences as a back-reaction of this perturbed thermal equilibrium could be summarised as
\begin{itemize}[leftmargin=*]
\item Both coronal heating and cooling processes are perturbed by any compressive perturbations (e.g. magnetoacoustic waves), causing a misbalance between heating and cooling rates.
\item The perturbed thermal equilibrium is important not only for traditional problems of the formation and evolution of prominences and coronal rain, but also for modelling and interpretation of MHD waves in the corona.
\item For typical coronal conditions, the characteristic timescales of thermal misbalance are demonstrated to be about observed oscillation periods of slow-mode waves, i.e. about several minutes.
\item The back-reaction that the wave experiences from the phenomenon of thermal misbalance includes dispersion (not connected with the waveguiding or cutoff effects traditionally considered in the corona) and enhanced frequency-dependent damping or amplification (via the energy exchange between the plasma and the wave).
\item The stability of the coronal plasma structures is sensitive to the heating model and the wavelength of compressive perturbations. Requiring a long-lived (stable) corona, one can use this effect for probing heating functions in various coronal structures.
For example, for a specific set of coronal plasma parameters, the RTV heating models were shown to be unstable to the perturbations longer than $\sim$100\,Mm.
\item The misbalance-caused dispersion of slow magnetoacoustic waves is most pronounced in the longer-period part of the spectrum, leading to the modification of characteristic sound and tube speeds. In some regimes of misbalance, this may lead to the formation of long-period slow-propagating wave trains from initially aperiodic, broadband perturbations. 
\end{itemize}

\noindent Below, we also list several potentially interesting and, in our opinion, promising avenues for future development of this research field
\begin{itemize}[leftmargin=*]
\item Given the importance of the thermal Field's length in the physics of solar prominences and coronal rain \cite{2020PPCF...62a4016A}, its generalisation for a non-constant coronal heating rate would allow for a fresh look at this problem, from both modelling and observational points of view. Likewise, practical implications of the acoustic Field's length for the dynamics of slow-mode waves in the corona are to be revealed.
\item Search for theoretically predicted slow-propagating quasi-periodic compressive wave trains in observations, using Refs.~\cite{2001A&A...377..691B, 2001A&A...380L..39B} as a starting point.
\item Development of a 2D theory of slow magnetoacoustic waves in coronal plasma loops with thermal misbalance, with a focus on the transverse fine structuring of the loop.
\item Revealing the effects and implications of thermal misbalance on other MHD eigenmodes of coronal plasma structures, such as fast magnetoacosutic waves and torsional Alfv\'en waves. The ground for this has been recently seeded in, for example, Ref.~\cite{2021SoPh..296...98B}, where nonlinear shear Alfv\'en waves in a plasma with thermal misbalance were considered.
\item Development and applications of the theory of thermal misbalance, accounting for the effects of parallel non-uniformity of the coronal plasma, i.e. $\rho(z)$, $T(z)$, and $B(z)$, and non-locality of the coronal heating function, ${\cal H}(\partial \rho/\partial z, \partial T/\partial z, \partial B/\partial z)$. An important question here is the possible modification of the acoustic cutoff frequency.
\item Accounting for the link between the coronal heating function and global parameters of the loop, such as loop length and transit time, and parameters of the photospheric driver (see e.g. Table 1 in \cite{2019ARA&A..57..157C}).
\item {An interesting future study could also address the effect of thermal misbalance on the nonlinear cascade and shock wave dynamics in the corona. In particular, the results obtained in Ref.~\cite{2018NatAs...2..951S} could be re-considered through the prism of formation of quasi-periodic trains of magnetoacoustic shocks in the regime of wave amplification caused by the misbalance \cite{2020PhRvE.101d3204Z, 2021PhFl...33g6110M}.}
\end{itemize}

\ack
The work was supported by the STFC consolidated grant ST/T000252/1 (DYK and VMN) and partly by the Ministry of Science and Higher Education of the Russian Federation by State assignment to educational and research institutions under Projects No. FSSS-2020-0014, 0023-2019-0003 (DIZ), and by Subsidy No. 075-GZ/C3569/278 (DYK). The work of DYK and VMN on the analysis of stability of slow waves and interpretation of results was supported by Russian Scientific Foundation grant 21-12-00195.

\section*{Data availability statement}
All data that support the findings of this study are included within the article (and references therein).

\section*{References}

\providecommand{\newblock}{}

\end{document}